\documentstyle[12pt,epsfig,subeqn,subeqnarray]{article}

\begin{document}

\title{Critical Examination of the Conceptual Foundations
of Classical Mechanics in the Light of Quantum Physics}
\author{Gennaro Auletta,
Institute of Philosophy, Universit\`{a} di Urbino
\\md0509@mclink.it}

\maketitle

\begin{abstract}
As it is well known, classical mechanics consists of several basic
features like determinism, reductionism, completeness of knowledge and
mechanicism. In this article the basic assumptions are discussed which
underlie those features. It is shown that these basic assumptions
--- though universally assumed up the beginnings of the XX
century  --- are far from being obvious. Finally it is shown that --- to a
certain extent --- there is
nothing wrong in assuming these basic postulates. Rather, the error lies in the
epistemological absolutization of the theory, which was considered as
a mirroring of Nature.

\end{abstract}

\section{Introduction}

Classical mechanics (CM) is one of the greatest achievements of human
knowledge. It is a compact building whose conceptual and mathematical
aspects have been known and studied in all details and consequences,
though the formation of the theory itself  was a difficult process which took
three centuries (XVII-XIX) to be completely achieved.

But, to the best of my knowledge,
some basic --- and sometimes only implicitly assumed ---
postulates of CM
have not been subjected to critical examination --- a state
of affairs which can be partly explained by the implicit
character of some of them. This situation, on the other hand, has as 
a consequence
that all basic postulates of the theory have not been considered in
their connection as a system --- contrarily to what happens for the formalism
of the theory, which from the beginning of XIX century is presented in
a systematic
form. In fact, to the best of my
knowledge, there is until today no handbook which introduces these
postulates in systematic order at the beginning of the
exposition\footnote{See \cite{kn:lelalif1976a}
\cite{kn:goldstein1} \cite{kn:dhestenes1986} \cite{kn:knud/hjor1995} 
for instance.}.

Such an enterprise is possible today because
quantum mechanics has shown, especially in the last twenty years,
that several assumptions of CM are far away from being above a
legitimate suspicion or a critical examination.

I will try to expose in a systematic form these
assumptions. In section \ref{sec-Determinism}, I shall examine CM's
determinism and the postulates from which it  stems. In section
\ref{sec-Reductionism} a similar analysis is devoted to
reductionism, while section \ref{sec-Mechanicism} shows that
mechanicism is equivalent to determinism plus reductionism.
Section \ref{sec-Completeness} examines
the postulate of the completeness of knowledge (and not of the
completeness of the theory itself, a problem which today cannot be
posed in this form). Sections \ref{sec-Classical Mechanics}--\ref{sec-Classical
Philosophy} are devoted to the  more general consequences of these
assumptions while in section \ref{sec-Conclusions} some
concluding remarks will follow.

\section{Determinism}
\label{sec-Determinism}

Everybody admits that CM is deterministic. But determinism is a more
complex assumption which supposes other, more basic postulates or
principles.

\subsection{Omnimoda determinatio}
\label{subsec-Omnimoda determinatio}

The most basic assumption of CM is the postulate of perfect
determination which may be expressed as follows: all
properties  of a physical system are perfectly determined, where a
{\it physical system} can be roughly understood as
an object or a collection of objects (somehow
interrelated) which can be (directly or indirectly) experienced,
and a {\it property} as the value which can be
assigned to a physical variable or observable. {\it Perfectly
determined} means that each variable of the system has at all times a 
definite value.

This assumption was always implicitly made. For all `classical'
physicists it was so self-evident that it was even not worth of mention.
In fact modern physicists --- like Galilei, Newton, and many
others --- inherited such assumption from all
past philosophy: Democritus, Plato, Aristotle, the middle-age 
philosophers until
the modernity never thought differently but all assumed that all
properties of being  are determined (the problem was obviously to
determine what the main or true being is). In fact Kant says: all
what exists is completely determined \cite[85]{kn:ikant1763} \cite[B
599--602]{kn:kant1}, which means that, between every possible predicate
of an object and its negation, one of the pair must be actualized.
Since every physical property can be reduced in a binary form (i.e.
using what in quantum mechanics are called projectors) --- for example
`moving at the speed of light' or `not moving at the speed of light'
(in some space--time context and by reference to a system, both to be
specified) ---, then the philosophical definition is a generalization of
the physical one and, so far as physical objects are considered, they can
been taken to be equivalent. For this examination Kant is
an especially interesting example because it has been often told that
his philosophy is strictly related with Newtonian mechanics. The
latin traditional expression for the complete determination is
{\it omnimoda determinatio} and can be found for example in \cite[par.
148]{kn:abaumgarten1739}.\footnote{On the problem see
also Leibniz's letter to de
Volder of april 1702 in \cite[239]{kn:leibniz2}.}

Now it turns out that quantum mechanics violates the {\it omnimoda
determinatio} at least through the superposition principle\footnote{For
argumentation of this
point see chapters 24, 30 and 46 of \cite{kn:gauletta2000}.}: In fact
if a quantum system can be in a superposition state, say of wave localized
in an arbitrary
region 1 and of another in an arbitrary but different
region 2, then, following Kant, it is certainly impossible to say
`either it is in
the region 1 or it is not' or `either it is in the region 2 or it
is not', or, more simply, `either it is in the region 1 or it is in
the region 2', i.e. quantum mechanics does not acknowledges a
generalized exclusion disjunction\footnote{On this specific point see
\cite{kn:hardergree}.}. There are no means to reduce such an
indetermination to subjective ignorance, so that it must be taken as
an ontologically basic feature of the theory.

\subsection{Continuity}
\label{sec-Continuity}

The {\it omnimoda determinatio} may be easily confused with determinism.
But they are not equivalent: In fact one can conceive a `classical' world
where every
`state of the world' (roughly: the complex of all states of all systems
within at a given instant) is in itself perfectly determined though
without relationship with past and future states, in the sense that the
world can jump from a state to another randomly. If this would be the
case, nobody could speak of a deterministic evolution (for a more formal
definition of determinism see next subsection).

Therefore, in order to obtain determinism we also need continuity.
This assumption is known as the principle of continuity and it states
that the variables associated to a physical system are supposed to be
continuous, which in physics means that, given two arbitrary values of
a physical variable, all intermediate possible real values are also
allowed.

The principle of continuity, though often not explicitly stated in
handbooks, was widely used already from the early days of CM: one may
remember here, for example, the use of the principle made by Galilei by
the law of fall\footnote{See \cite[130--131, 181]{kn:mach1}
\cite[223]{kn:mach2}.}.

Obviously the point of major conflict between CM and quantum
mechanics was continuity, which is rejected by the quantum
postulate (the values of quantum observables can be discontinuous)
and by the fact that quantum systems can jump\footnote{The quantum
postulate was assumed by Planck \cite{kn:planck1, kn:planck2} and
generalized by Bohr \cite{kn:bohr3}. About the formalism of quantum jumps see
\cite{kn:carmichael1}.}.

\subsection{Determinism}
\label{subsec-Determinism}

Sufficient and necessary conditions of determinism are the postulate
of perfect determination and the principle of continuity.
In fact, given the perfect determination of the state of a system at
an arbitrary initial state, if its dynamic variables (for example
the energy) are continuous, then also every future state of the system
will be perfectly determined and unique, i.e. without
alternatives or without branching\footnote{One may
think at Laplace's dictum that nothing is uncertain: see
\cite[134--35]{kn:cassirer3}
\cite[7]{kn:jearman1}. The latter book represent a good analysis of
classical determinism.}. As it is known, a mathematical
formulation of this assumption is given by Hamilton's canonical
equations, which in terms of Poisson brackets can be written
\begin{eqnarray}
\dot{q} = \{H, {q}\}, &  & \dot{p} = \{H, {p}\}.
\label{eq: poisson}
\end{eqnarray}
The Hamiltonian $H$, the energy function of the system,
fully describes the system and all its future (and past)
evolution, and it is expressed
in terms of the position $q$ and the momentum $p$. Note, however, that
determinism is not the same as predictability: in fact it is well
known that, for a large class of problems, almost identical initial
conditions can have very different solutions for later times [see
also subsection \ref{subsec-Linearity}].

Determinism too is an ancient philosophical concept, introduced
probably by ancient atomism and further developed by ancient stoicism.

In quantum mechanics the Schr\"{o}dinger equation is also deterministic,
but here what evolves deterministically is an amplitude, i.e. it
is  --- paradoxically --- a determinism which is intrinsically probabilistic.
In fact, in the general
case, we cannot foresee what values the observables will have; we can only
write their probability distributions.
As we have already said, the break of continuity is a major feature
of quantum mechanics. For this reason Bohr \cite{kn:bohr5}
\cite{kn:bohr6} spoke of a break of determinism and of causality as such.
Causality may be here understood in a strict and in a wide sense,
following the distinction between determinism and predictability. In
fact, in a
strict sense it may be taken as equivalent to determinism. In a wide
sense it may be understood as predictability, and then it
should be rather taken as equivalent to mechanicism [see section
\ref{sec-Mechanicism}] to the extent in which also linearity and
separability are necessary in order to have a predictable future.

One could think that, behind perfect determination and
continuity, determinism also supposes conservation laws. We may
understand the problem of conservation in two forms:
\begin{description}
	\item[i)]  nothing which is physical can disappear;
	\item[ii)] in an isolated system certain
physical quantities such as the angular momentum are conserved.
	\end{description}
On the
second point we shall return later [see subsection \ref{subsec-Isolability}].
Point i) is a general
statement about the conservation of energy (which obviously is a
conserved quantity also in the second  sense). In fact energy is the basic
physical quantity which, following our physics,
can be transformed but never destroyed, a
fact that is expressed in a general form by Einstein's equation
\begin{equation}
E = mc^{2},
\label{eq:einstein'seq}
\end{equation}
which is valid also in the case of annihilation of
particle--antiparticle pairs (in fact, as it is known,
their mass is transformed in the energy of `outgoing' photons).

In the statement i) it is the universe as a whole which is considered
as a closed system. It is evident that this statement is a more basic
one --- but also a weaker one --- than statement ii).
In fact, we could think a world where there can be no strict
conservation of the energy in the sense of statement ii) and notwithstanding
would be deterministic. For instance, there could be an universal but
unknown and unknowable `ether' such that all observable physical
system lose part of their energy. Then the energy would be conserved in the
sense of statement i) because a form of energy is transformed in
another form by the action of ether, but it is not conserved in the
sense of statement ii) because, for example, we could have a
a physical law of this form:
\begin{equation}
\dot{p} = F - \alpha p,
\label{eq:noconserv}
\end{equation}
where $F$ is the force and $\alpha$ some parameter.
Obviously the `path' or the `trajectory' of the every physical system
should be always calculable, i.e. the loss of energy should follow
strict laws and not be random. Otherwise the world could be not
deterministic. On the other hand, as we have said, even if the 
`physics' in this
universe should be expressed in terms of equations like
(\ref{eq:noconserv}), the ether is not something which is outside of
such an universe, so that, in a certain objective (or meta--physical)
sense (God's point of
view?), proposition i) is satisfied. But the difficulty of this
position is to admit the existence of something physical which in principle
cannot be experienced.

\section{Reductionism}
\label{sec-Reductionism}

Reductionism, as we shall see in that what follows, is another basic
piece of CM and, as determinism, it supposes other assumptions which
need to be preliminarily examined.

\subsection{Materialism}
\label{subsec-Materialism}

One may wonder that the assumption of materialism is basic for
CM since one may think that it is a metaphysical assumption without
consequence or relevance upon a physical science as CM is. But this is
not  the case: CM is a mechanics, i.e. a theory of the motion of
bodies and of the forces which act upon them. And a body is
necessarily a material entity.

In fact the existence and the basic
properties of matter were assumed and defined from the beginnings
already by Galilei\footnote{See \cite[II, 387--89]{kn:cassirer1}. See
also \cite[248--49]{kn:mach1} \cite[106--107]{kn:hall1}.}
and by Newton. In the third {\it regula philosophandi} of book III of
the {\it Principia} \cite[552--55]{kn:newton1}, Newton makes a catalogue of
properties of matter (bodies): Extension (a cartesian property),
hardness, impenetrability, capacity to move, inertia\footnote{See
\cite[ch. 7]{kn:koyre1} for commentary.}. About hardness, in
\cite[388--92]{kn:newton2} it is explained that the parts of all
homogeneous hard bodies which fully touch one another stick together
very strongly. From their cohesion Newton inferred that particles
attract one another by some force, which in immediate contact is
exceedingly strong. On the other hand all bodies seem to be composed of
hard particles; for otherwise fluids, as water, would not freeze, or
fluids as ``spirit of nitre and mercury'' would not become hard
``by dissolving the mercury and evaporating the flegma''. And therefore
hardness can be reckoned as the property of all uncompounded matter.
So far Newton. It
is then evident that all fluids can be reduced to hard bodies by
freezing or evaporating: In this case the particles cohere fully,
which in turn means that some bodies are not hard only because they
are to a certain extent rarefied, i.e. there is some vacuum between
the particles\footnote{An important difference with respect to
Descartes \cite[33--34, 1205--110]{kn:koyre3}.}.
In other words, following Newton, all matter can be
reduced to some ground `state' in which it is fully homogeneous and
hence inelastic. In fact elasticity is possible only if there is some
internal structure to the matter, which is excluded by the postulated
homogeneity.

One may discuss --- and Newton himself had no final
position about\footnote{See the mentioned regula III.} --- if matter is
a continuous medium divisible {\it in
infinitum} or it is composed by elementary corpuscles which are strongly
bound and fixed together by adhering to each other.
However, the consequence is that, by full homogeneity and/or rigidity,
in case of collision of two bodies moving at the
same speed from opposite directions, they will coalesce at the point
of collision (because fully inelastic).\footnote{For examination see
\cite[ch. 9]{kn:koyre1}.} One may say that the kinetic energy has been
transformed in some activity of the particles composing the body, but
precisely this is impossible because there is no internal structure and
no possibility of the particles to translate, to rotate or vibrate
relatively to one another\footnote{For all the problem of bodies'
collision see \cite[310--31]{kn:mach1}.}. In a general way note that
Newton had not included the force as an intrinsic
property of matter as such --- i.e. forces can only act
`from outside' upon the matter. In fact Newton only attributes a {\it vis
inerti\ae} to the matter and  says \cite[397--98]{kn:newton2} that it
``is a passive
principle by which bodies persist in their motion or rest, receive motion
in proportion to the force impressing it, and resist as much as they are
resisted. By this principle alone there never could have been any motion
in the world. Some other principle was necessary for putting bodies into
motion; and now they are in motion, some other principle is necessary for
conserving the motion. For from the various composition of two motions,
'tis very certain
that there is not always the same quantity of motion in the world.
[\ldots] it appears that motion may be got or lost. But by reason of the
tenacity of the fluids, and attrition of their parts, and the weakness of
elasticity in solids, motion is much more apt to be lost than got, and is
always upon the decay. For bodies which are either absolutely hard, or so
oft as to be void of  elasticity, will not  rebound from one other.
Impenetrability makes them only stop. If two equal bodies meet directly
{\it in vacuo}, they will by the laws of motion stop where they meet, and
lose all their motion, and remain in rest, unless they be elastic, and
receive new motion from their spring.'' Therefore Newton  concludes
\cite[401--402]{kn:newton2} that it
seems to him that ``these particles [of matter] have not only a {\it vis
inerti\ae} \ldots but also that they are moved by certain active
principles, such as is that of gravity, and that which causes
fermentation, and the cohesion of bodies.'' As it is clear in the following
pages of the {\it Optics} and in other places, these principles are due
to the  direct action of God. Therefore, one can understand that Leibniz,
in his letter to the princess of Wales \cite[VII, 352]{kn:leibniz2}, defend
the conservation law of 'force and energy' against Newton. And it is
interesting that, in his first answer, Clark  writes
\cite[VII, 354]{kn:newton2} that God ``not only
composes or puts things together, but is himself the Author and continual
Preserver of their original forces and moving powers''.

Therefore we see that the materialism assumed since the early days of
CM is far from being obvious, and the idea of a fully homogeneous
matter was  very soon abandoned.
In quantum mechanics there can be no question
of perfectly hard and localized corpuscles: To quantum
entities is intrinsic a wave-like behavior or some fuzziness. Therefore it is
better to speak of extended particles\footnote{On this point see
chapters chapters 30 and 33 of \cite{kn:gauletta2000}.}. On the other
hand properties as the hardness or impenetrability seem inadequate to
microentities as we know them now.

\subsection{Linearity}
\label{subsec-Linearity}

Linearity is an important property of classical systems. In itself it
is essentially a mathematical property, because it consists in the
requirement that the basic equations of CM must be linear, i.e.
reducible to a form like
\begin{equation}
a_{0}(x)y^{(n)} + a_{1}(x)y^{(n - 1)} + \ldots + a_{n}(x)y = f(x),
\label{eq:lineareq}
\end{equation}
where $a_{0}(x), a_{1}(x), \ldots, a_{n}(x)$ are coefficients,
$f(x)$ is some function and $y^{(n)} $ the $n$--th derivative of $y$.
But linearity has
a conceptual relevance to the  extent in which it excludes
feed--back, i.e. self--increasing processes.

It is linearity which allows an important aspect of the
`reductionistic methodology' of CM: the factorization between
component `elements' of a system, for example the decomposition of
motion in components by Galilei, the decomposition of forces by
Newton or the decomposition of harmonic components\footnote{For these
examples see \cite[144--45, 191--92]{kn:mach1}.}. In other words if
the cause (the force) ${\rm C}_{1}$ produces the effect (the
acceleration) ${\rm E}_{1}$ and the cause (the force) ${\rm C}_{2}$ the
effect (the
acceleration) ${\rm E}_{2}$, then ${\rm C}_{1} + {\rm C}_{2}$ produces
${\rm E}_{1} + {\rm E}_{2}$. This principle is often called principle of
(classical) superposition.

One could think that in CM
a small perturbation on a given system or the weak interaction
of this with another system only causes a small deviation in
the trajectory of the system in the phase space, such that, normally,
the system will `absorb' it and return on the ancient deterministic
path. But a perfect classical system can show such a dependence on
initial condition that its evolution can be chaotic (in fact in
chaotic regime this dependence is expressed by a strong divergence of
initial very close, indistinguishable trajectories in phase space).
Note that, in the chaos theory, chaos itself
is intrinsic and deterministic and not  stochastic and
extrinsic --- in other words it is not due to random fluctuations of the
environment or to noise\footnote{On the
point see \cite{kn:hgschuster1988} \cite{kn:druelle1989}.}.
In fact there can be chaos also by Hamiltonian systems.

Linearity is not violated by quantum mechanics. In fact
Schr\"{o}dinger equation is linear, and any attempt to introduce
non--linear terms in this equation has up to now
failed\footnote{A non--linear equation for quantum mechanics was
proposed in \cite{kn:bibim1}. Shimony proposed an experiment
aiming to verify if there are non--linear terms and if they have the
magnitude proposed by Bialynicki--Birula and Mycielski
\cite{kn:shimony5}. A later experiment performed on these outlines
tendentially excludes such terms \cite{kn:shuatarh1}. Obviously
this does not mean that the methods of quantum mechanics and chaos theory
cannot be combined. They can be, and actually are unified in what is 
today known as
`quantum chaos'.}.

\subsection{Separability}
\label{subsec-Separability}

Separability is another key feature of CM. But it is again an
implicit assumption and firstly in 1935, as CM was confronted
with quantum mechanics, it was stated explicitly by Einstein and
co--workers \cite{kn:epr1}. The principle of separability may be
expressed in the following way:
given two non--interacting physical systems, all their physical properties are
separately determined, or, in other terms, the result of a measurement
on one system cannot depend on a measurement performed on
the other system. The meaning of the principle is the following: two
systems can be interdependent only through a physical interaction (for
example some form of potential energy).

Again quantum mechanics violates the separability principle by a
consequence of the superposition principle for multiparticle systems:
entanglement. In fact for entangled subsystems, it is not possible to
factorize the probabilities of the outcomes of experiments performed
on each subsystem locally. In other words, probabilities calculated on
one of two `distant' subsystems, even if they do not
physically interact, are not independent\footnote{There exists a  wide
literature on this subject. For a summary see chapters 31 and 34--35 of
\cite{kn:gauletta2000}.}.

\subsection{Reductionism}
\label{subsec-Reductionism}

Now we may summarize the results of this section by saying that
materialism plus linearity plus separability are sufficient and
necessary conditions of reductionism.
Roughly speaking, by
reductionism it is usually meant that a system is given as the ``sum''
of its constituent components, or, equivalently, that any system can
be divided into ``elementary'' parts. The aim of reductionism is then
to find the ultimate elements of matter which cannot be further
reduced. To our knowledge there is no certainty (and even
doubts) that such a task will ever be accomplished. One speaks today,
for example, of quarks and leptons as `divisible' particles.
However, quantum mechanics violates this type of reductionism because 
it violates
the separability principle --- and does not, as we have seen,
violates linearity (leaving aside the problem of materialism). In
fact it is evident that, if separability is violated, no reduction of a
whole to `parts' is possible because the parts could be not  treated as
independent systems.

On the other hand, reductionism can be also understood as the reduction of
more complex theories and sciences as chemistry and biology to physics
and especially to quantum mechanics (this may be called {\it epistemological}
or {\it methodological}
reductionism relatively to the first type, which may be called {\it
ontological} reductionism). It is true that quantum
mechanics shows its effects (entanglement, for example) also at
mesoscopic level. But this means anyway that the mesoscopic or the
macroscopic world are only `illusions', apparent realities. In fact
the process of decoherence and especially of localization which goes
together  with decoherence, especially when the number and the complexity
of systems grows, is throughout objective\footnote{See chapters 17 and
24--25 of \cite{kn:gauletta2000}.}. On the other hand, no necessity
arises to conceive of methodological reductionism as a one--way
operation: If one speaks of reduction to more elementary objects, one
should speak --- with more  reason --- of a methodological reduction of
microscopic equations for the constituents of a system (via coarse graining)
to differential equations for macroscopic variables, and from these
(via numerical calculations of Poincar\'{e} sections) to low
dimensional Poincar\'{e} maps\footnote{See
\cite[14--16]{kn:hgschuster1988} \cite[63--78]{kn:berge/pomeau/vidal1984}.}.

\section{Mechanicism}
\label{sec-Mechanicism}

Sufficient and necessary conditions of mechanicism are determinism and
reductionism. No `classical' mechanics can violate the one or the other.
In facts mechanicism consists in the theory that, given an input
(some force) we have a fully automatic and proportional output
(some acceleration), which
would be surely impossible if the whole system were more than the
sum of the `parts' (i.e. if the requirement of reductionism would
be violated), or if it would show a random reaction to a given action
(i.e. if the requirement of determinism would be violated). On the
other hand, a system satisfying the features of determinism and
reductionism would be necessarily mechanic. In fact we distinguish the
behavior of organic life from a pure mechanical behavior exactly
through the violation of the one or of the other requirement or of both.

\section{Completeness}
\label{sec-Completeness}

The possibility of a complete knowledge in CM is dependent on other 
assumptions,
namely determinism and isolability. Let us examine firstly the
assumption of isolability.

\subsection{Isolability}
\label{subsec-Isolability}

CM assumes that isolated systems are possible; i.e. that we can
always theoretically treat and experimentally (at least in principle)
generate a system without physical interdependence with other systems
or with the environment. It is the isolability which guarantees
conservation laws of pertinent quantities. In fact angular momentum,
energy or motion can be conserved only if the system is considered as
isolated from others, i.e. there is no interaction such as to cause
dispersion or no action of an external force such as to change
its motion.
Quantum mechanics does not apparently violate this assumption. But it
may be asked if there are actually isolated quantum--mechanical
systems and even more if macroscopic systems can be fully isolated.

\subsection{Completeness}
\label{subsec-Completeness}

In CM it is supposed that one can perfectly know (at least in
principle) all properties of a given system. In other words the
properties of the object system can be perfectly
measured. Therefore it is postulated that the measurement
errors  can be --- at least in principle ---
always reduced below an arbitrarily small quantity
$\epsilon$. Hence this assumption may be
called the postulate of reduction to zero of the measurement
error.

Note that this postulate is not a direct consequence
of the principle of perfect determination only, because it can be the case
that a system is objectively but not subjectively perfectly determined.
It supposes continuity too:
in fact if the pertinent variables were
discontinuous, then we could not approximate to a point-like
value in a given interval. Hence it presupposes determinism (which,
as we know, is
equivalent to perfect determination plus continuity).
But isolability too: In fact if the system could never be really
isolated, we could  never know its properties perfectly, even not in
a very large time interval, because, during the flow of time, it may
be that small interactions with external systems cause small
uncertainties in the measurement results so that --- even if these 
uncertainties
do not cumulate --- one cannot go beyond a certain threshold.

If we speak of the perfect knowledge of all properties of a given
system at the same time, then this assumption is obviously violated in
quantum mechanics through the uncertainty principle. In fact this
principle states that, by increasing the knowledge or the
determination of an observable of a conjugate pair, the complementary
observable must proportionally increase its uncertainty.

\section{Classical Mechanics}
\label{sec-Classical Mechanics}

We can now draw the first general conclusion from the above analysis:
CM consists of both mechanicism and completeness (of knowledge).
There is no doubt that there can be no CM without mechanicism.
But one may think that completeness is not a necessary condition of CM.
This is not the case because CM is actually so built that
a perfect transparency of the
object system to the knowledge corresponds to the perfect
ontological determination of it.
But it could also be not otherwise: For a physicist
the primary questions are objective and not subjective: In order to
admit an incomplete knowledge together with the assumption of
mechanicism --- and hence of a perfect ontological determination ---, one
should know some basic limitations of human mind, which in
principle exclude the possibility for human beings
of perfectly knowing systems which are objectively
perfectly determined. But no such problems have
ever been found.

\section{Classical Gnoseology}
\label{sec-Classical Gnoseology}

Classical mechanics has been developed together with what may be called a {\it
classical} gnoseology --- i. e. the gnoseology of Galilei, Spinoza, Newton,
Kant and many others. Classical gnoseology certainly supposes the
completeness of knowledge, i.e. that the properties of being can be
perfectly known. But it also supposes what may be called a `mirroring'
theory. Explicitly: classical gnoseology considers the act of
knowledge as a mirroring of the properties of the object.

In other words, knowledge is understood as a reproduction of objective
and given data and
not as a form of interaction between subject and object. This
understanding of knowledge is very ancient and can also be found by
philosophers as Plato. Several philosophy schools have shared this
pointof view. Obviously, there is no agreement between several schools
about what is the being to be reproduced (ideas as platonic
substances, atoms, forms, material objects, and so on). When knowledge is so
understod, then one assigns to the subject a mere reproductive and
representative role.

However this view is no so evident. In fact pragmatism\footnote{On this point
see \cite{kn:cpeirce1878a, kn:cpeirce1878c}.} proposes a different theory
of knowledge. It is seen as a
problems--solving enterprise which, by starting with a problem,
assumes a hypothesis (under many others possible ones) because it can
solve in a satisfactory manner the conflicts or the contradictions
arisen from the problem itself. This is not the place where to
examine this subject in details, but I think that this explanation of how
theories work and are generated is far more satisfactory for
describing scientific knowledge than the
traditional, classical approach. I only wish to stress following
aspects of this explanation:
\begin{description}
	\item[i)] Subject and object are not understood as static beings and
	knowledge not as a form of translation of data into a mind (and how
	would it be possible?).
	\item[ii)] Experience is dynamic and comprehends `subject' and
	`object'\footnote{On this point see \cite{kn:dewey1}.}.
	\item[iii)] Knowledge is open and never represents a final answer.
	\item[iv)] Knowledge is a form of praxis and the theory is not
	completely separated from other human activities.
	\end{description}

\section{Classical Philosophy}
\label{sec-Classical Philosophy}

Classical philosophy, the main stream in XVII-XVIII centuries, is
compound of CM plus classical gnoseology. That
philosophers and physicists of that age have acknowledged all or almost
all the above principles can be seen from the following examples.
Let us first take
Kant's examination of the ontological proof of the existence of
God\footnote{In \cite[B 627]{kn:kant1}.}.
Kant says
that when I affirm that {\it God exists}, I add no new predicate to the
concept of God; rather I pose only the subject (God) in itself with all his
predicates, i.e. the object, in relationship with my concept. Both,
the object and the concept, must contain the same. In other words, in
Kant's terminology, what
is real does not contain something more than what is only possible
(the concept). If the object should contain more than the concept, then
the latter will not express the whole object and will therefore be
not adequate to this object. So far Kant. In this argumentation, the
{\it ominimoda determinatio} is always taken for granted and
three additional principles are (implicitly) assumed: that the
concept is isomorphic with the object (the predicates contained in the
concept corresponds to properties that the object
has: it is the mirroring theory);
that therefore an adequate knowledge must be
complete (all properties of the object must be considered in the
concept); and finally that one can consider the object
`in itself', i.e. in complete isolation from other
objects (it is the assumption of isolation). Since this is a general
arguments which goes beyond to the
specific problem of the existence of God, one can consider
the object without relationship with the
other objects of the universe. It is true that in \cite[B
599--602]{kn:kant1} one speaks of the {\it omnimoda determinatio} as
an ideal, but in the above proof it is taken as an ontological fact.

To my knowledge Kant never rejected the continuity and perhaps he had
nothing against linearity. He surely
assumed a form of materialism: Since our knowledge can only happen in
an experience which is intrinsic spatio--temporal \cite[B 33--73]{kn:kant1},
then the objects
of knowledge can only be bodies; and in fact Kant discusses the problem
of matter \cite[B 230, 277--78]{kn:kant1} and excludes that
the subject of knowledge can also be object of
knowledge \cite[B 152--165]{kn:kant1}. Of all above postulates
only separability remains; but, as
already said, it has been the object of scientific analysis only in
the 1930s.

Then let us also briefly discuss the assumptions (but not the details
of the
argumentation) of the article of Einstein and
co--workers \cite{kn:epr1}. There is no doubt that it acknowledges
the {\it omnimoda determinatio}. In fact the aim of the article is to
show that there can be elements of the reality which cannot be
represented in quantum mechanics due to its `uncertain' character
(uncertainty principle) --- in fact, as it is well known, Einstein
thought that quantum mechanics could only represent a statistical
(and therefore incomplete) theory of microentities.
Specifically, the aim of the article is to show that quantum
mechanics violates a sufficient condition of reality, which may be
expressed as follows: If, without in any way disturbing a system,
we can predict with
probability equal to unity the value of a physical quantity,
then there exists an element of the physical reality corresponding
to this physical quantity. It is evident that two things are
supposed here: first that the reality is perfectly
determined in itself; second, that one can also know it perfectly (our
completeness condition).
Continuity is
evidently acknowledged in the formal development of the argument. So
there is no doubt that the article also acknowledged determinism (= {\it
omnimoda determinatio} + continuity).
Though no word is said about materialism and linearity, the core of
the article is represented by a strong defense of the principle of
separability (here for the first
time formulated), so that one can suppose that reductionism too was a
valid assumption for Einstein and co--workers.

But the article goes even further.
In fact two definitions are formulated with great emphasis at the beginning:
That of correctness and that of completeness.
It is said that a theory is totally correct if
every element of the theory has a counterpart in reality:
In other words a totally correct theory is one without
superfluous theoretical terms. It is evident that the necessary
condition for assuming this definition is the mirroring
theory: If theories could not mirror reality, could also not mirror
reality correctly. About completeness it is said that
a theory is complete if every element of reality has a counterpart in it
--- it is evident that correctness together with completeness  establish an
equivalence relationship between physical theory and reality. This definition
of the completeness of a theory is much stronger than that previously 
formulated.
In conclusion CM and classical gnoseology, and therefore
classical philosophy as such, are defended in Einstein's article.

It is very interesting that Kant and Einstein --- both scientists and
philosophers --- defend essentially the body of classical philosophy,
and that the latter does it in open conflict with quantum
mechanics.

\section{Conclusions}
\label{sec-Conclusions}

Summing up, CM can be schematically represented as in the figure.

\begin{figure}[htpb]
\centerline{\hbox{\epsfig{file=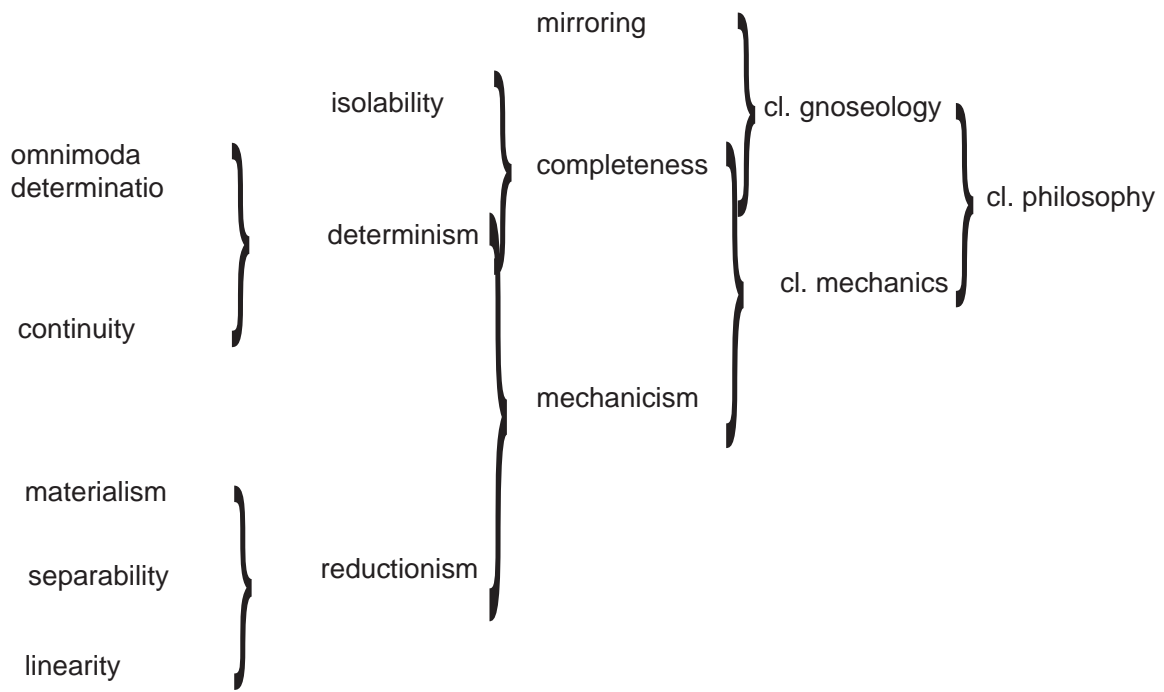,width=6in}}}
\caption{\small
Schematic representation of the basic postulates and principles of CM.}
\label{fig:CMb}
\end{figure}

CM has been for three centuries the model of what Science is and
should be. Then it is a little surprising that its basic assumptions
were assumed without critical examination. But two points are here 
very important:

\begin{description}
	\item[i)] Without quantum mechanics and its consequences 
nobody would have
perceived the problems hidden in assumptions which ultimately stem
from the common sense or from a refinement of the ordinary experience
about macroscopic objects. This does not mean at all that this experience
is in itself wrong. We live and act in a macroscopic world where the
struggle for life is the most important thing, and for this practical
purpose it makes no sense --- and it is perhaps even dangerous --- to
assume, for
instance, that objects are partly not perfectly determined or
fuzzy\footnote{And, with high probability, also macroscopic objects are partly
fuzzy; in fact one has shown theoretically and finally experimentally
that, at mesoscopic level, `Schr\"{o}dinger
cats' are possible \cite{kn:momekiwi1} \cite{kn:bruhadremam1}.}
	\item[ii)] But neither CM's assumptions are wrong as such. CM 
has been in fact a
powerful tool in order to explore Nature and establish some basic
features of the physical world. Stated in other terms, for all that
one knew at that time, CM worked --- and still works ---
very well. What is wrong is only the supposition
that CM's assumptions and laws are objective in the sense that they
mirror what Nature is in itself. In other words, what was and is wrong
about CM is a `mirroring' gnoseology and epistemology which has
produced an absolutisation of the this physical theory. In other
words, we have here a confirmation {\it e contrario} of the
rightness of the point of view of pragmatism.
	\end{description}

\section*{Acknowledgments}

This article is born from a synthesis of several lessons about Classical
Mechanics and discussions with my
students in the Gregoriana University in Rome.

The present article owes very much to
Prof. Giorgio Parisi whose deep insights were for me especially
enlightening.
I also thank Dr. Mauro Fortunato and Dr. Valeria Mosini
who red the manuscript and gave to me plenty suggestions.

\end{document}